\documentclass[11pt]{article}
\usepackage{amssymb}
\usepackage{amsfonts}
\usepackage{amsmath}
\usepackage{graphicx}

\begin{document}

\date{}
\title{Quintessential Cosmological Scenarios in the Relativistic Theory of
Gravitation\thanks{%
To appear in Proc. XXV Int. Workshop on Fundamental Problems of High Energy
Physics and Field Theory (25-28 June 2002, Protvino, Russia)}}
\author{V. L. Kalashnikov \\
Technical University of Vienna, \\
Gusshausstr. 27/387, A-1040, Vienna, \\
fax: +43-1-5880138799, \\
e-mail: v.kalashnikov@tuwien.ac.at}
\maketitle

\begin{abstract}
It is shown that the accelerated expansion of the universe in the framework
of the relativistic theory of gravitation can be achieved by the
introduction of the quintessential term in the energy-momentum tensor. The
value of the minimum scaling factor and the modern observational data for
the density and state parameters of the matter give the rough estimations
for the maximum graviton mass and the maximum scaling factor. The former can
be very low in the case of the primordial inflation and the latter can be
extremely large for the scalar field model of the quintessence. In any case,
the massive graviton stops the second inflation and provide the closed
cosmological scenario in the agreement with the causality principle inherent
to the theory.
\end{abstract}

\section{Introduction}

The relativistic theory of gravitation (RTG) \cite%
{logunov1,logunov2,logunov3} disagrees with the Einstein's general
relativity (GR) in the crucial point: it denies the total geometrization and
considers the gravitation on the basis of the classical Faraday-Maxwell's
field approach. This means that there is a topologically simple background
spacetime of Minkowski type, which can be restored in any situation. As a
result, we can detach the physical content from an arbitrary geometrical
game with co-ordinates. This converts the gravitation from the
tensor-geometrical concept to the tensor-field one and puts it on the
unified level with another fields.

Formally, the RTG can be considered as the bi-metric theory of gravitation
\cite{rosen1,rosen2}. However, in the RTG the effective Riemannian spacetime
produced by the gravitational field is essentially separated from the
Minkowski background because the latter is presented in the field equations
(see next section). Naturally, this transforms the solutions of the field
equations and has the pronounced physical consequences. For example, the
singularity disappears and the graviton acquires the nonzero mass.
Nevertheless, the basic observational consequences of the RTG coincide with
those in the GR (for instance, Mercury perihelion motion, time delay and
spectral shift in the gravitational field, see \cite{logunov2}).

The application of the RTG for the cosmology produces some astonishing
results, viz., in virtue of the field equations the
Friedmann-Robertson-Walker cosmology admits only the flat global efficient
Riemannian spacetime without initial singularity and with oscillating time
behavior \cite{logunov2,logunov3}. The initial cosmological expansion is
stimulated by the antigravitation, which is caused by the massive gravitons
in the strong gravitational fields. The initial temperature is defined by
the graviton mass and can be too low to create the undesirable relics (e.g.
monopoles). So, the problems of the cosmological spacetime flatness, the
source of the initial expansion, the cosmological singularity and the
absence of the relics find in the RTG a natural solution. However, in this
theory there are some disagreements with the modern observational data. As
it is known, the latter suggests the accelerated expansion of the universe
at present (see, for example, \cite{riess,permutter}). But in the RTG the
accelerated expansion is possible only during a very short stage of the
initial evolution and the subsequent expansion has a definitely decelerated
character.

As it is well known, the accelerated cosmological expansion in the framework
of the GR can be obtained \textquotedblleft by hand\textquotedblright\ due
to an insertion of the so-called cosmological constant in the field
equations (for a review see \cite{weinberg}). This constant can be
considered as a part of the geometrical structure of the GR because it is a
natural consequence of the variational principle \cite{lovelock}.
Alternatively, it is possible to treat the cosmological constant as the
vacuum zero-point energy. But in the both cases its value is too small and
can not be attributed to any known physical scale.\

The situation in the RTG is more complicated by virtue of the vacuum
stability principle: the absence of the material fields reduces the
effective Riemannian spacetime \ to the Minkowski one. Hence, the
cosmological constant can not be introduced by hand and is to have the
gravitational nature concerned with the nonzero graviton mass. As a result,
the cosmological constant-like action of the massive graviton in the RTG
produces the \textit{deceleration} of the cosmological expansion.

Nevertheless there exists an approach, which considers the accelerated
expansion of the universe as a manifestation of some matter possessing an
unusual state equation  $p=w\rho $ (where $p$\ is the pressure and $\rho $\
is the density). This matter usually is called as X-matter or \textit{%
quintessence}. If its state parameter $w$ lies between the limits of the
strong and week energy conditions (i.e. $-1\leq w\leq -\frac{1}{3}$), the
domination of such matter produces a repulsion causing the accelerated
expansion of the universe \cite{peebles}. The best candidate here is a
certain scalar field whose potential energy dominates at present (the survey
can be found in \cite{binetruy}, for example).

In this article we shall consider the implementation of this idea in the RTG
framework. As a result, some restrictions on the key parameter of the
theory, i. e. the graviton mass, will be obtained.

\section{Basic equations}

The field equations for the gravitational field in the framework of the RTG
are based on the assumption that the universal character of the gravitation
allows introducing the effective Riemannian spacetime \cite{logunov3}:

\begin{equation}
\widetilde{g}^{\mu \nu }=\widetilde{\gamma }^{\mu \nu }+\widetilde{\varrho }%
^{\mu \nu },  \label{eq1}
\end{equation}

\noindent where \ $\widetilde{g}^{\mu \nu }=\sqrt{-g}g^{\mu \nu },~%
\widetilde{\gamma }^{\mu \nu }=\sqrt{-\gamma }\gamma ^{\mu \nu },~\widetilde{%
\varrho }^{\mu \nu }=\sqrt{-\gamma }\varrho ^{\mu \nu }$ are the densities
of Riemannian metric tensor, Minkowski metric tensor and gravitational field
tensor, respectively. In this case the Lagrangian density for the
gravitational field is the function both $\widetilde{\varrho }^{\mu \nu }$\
and $\widetilde{\gamma }^{\mu \nu }$. It is essential that the effective
Riemannian spacetime is completely defined for the given Minkowski
co-ordinates, i. e. $g^{\mu \nu }$ is their single-valued function. Hence,
the topology of the effective spacetime is quite simple. Let us consider the
infinitesimal coordinate transformation induced by the translation vector $%
\zeta ^{\mu }$:

\begin{equation}
x^{\mu }{\acute{}}=x^{\mu }+\zeta ^{\mu }.  \label{eq2}
\end{equation}

\noindent Then the field-dependent metric density of the effective spacetime
is changed as:

\begin{equation}
\widetilde{g}^{\mu \nu }{\acute{}}=\widetilde{g}^{\mu \nu }+\delta _{\zeta }%
\widetilde{g}^{\mu \nu }+\zeta ^{\lambda }D_{\lambda }\widetilde{g}^{\mu \nu
},  \label{eq3}
\end{equation}

\bigskip \noindent \noindent $\delta _{\zeta }$ is the Lie variation and $%
D_{\lambda }$ is the covariant derivative on the Minkowski (i.e. background)
spacetime. If the Lagrangian density for the gravitational field depends
only on $\widetilde{g}^{\mu \nu }$ and its derivatives, then the
transformation (\ref{eq3}) changes this density only on a divergence. On the
basis of (\ref{eq3}) we can define the gauge group preserving the field
equations and the background metrics. Let's Eq. (\ref{eq3}) describes the
transformation induced by the infinite-dimensional gauge group with the
gauge vector $\zeta ^{\mu }$. In contrast to\ the coordinate transformation,
this gauge transformation does not effect the background: $\delta _{\zeta }%
\widetilde{g}^{\mu \nu }=\delta _{\zeta }\widetilde{\varrho }^{\mu \nu }$.

The simplest Lagrangian density, which is changed only on a divergence by
this gauge transformation, can be constructed from $\sqrt{-g}$ and $%
\widetilde{\Re }=\sqrt{-g}\Re $ ($\Re $ is the scalar curvature of the
effective spacetime). Let us define (see \cite{logunov2,logunov3}) the
scalar curvature density through the tensor $F_{\nu \lambda }^{\mu }$:

\begin{equation}
F_{\nu \lambda }^{\mu }=\frac{1}{2}g^{\mu \kappa }\left( D_{\nu }g_{\kappa
\lambda }+D_{\lambda }g_{\kappa \nu }-D_{\kappa }g_{\nu \lambda }\right) .
\label{eq4}
\end{equation}

\noindent Then

\begin{equation}
\widetilde{\Re }=-\widetilde{g}^{\mu \nu }\left( F_{\mu \nu }^{\lambda
}F_{\lambda \kappa }^{\kappa }-F_{\mu \kappa }^{\lambda }F_{\nu \lambda
}^{\kappa }\right) -D_{\nu }\left( \widetilde{g}^{\mu \nu }F_{\mu \kappa
}^{\kappa }-\widetilde{g}^{\mu \kappa }F_{\mu \kappa }^{\nu }\right) .
\label{eq5}
\end{equation}

\noindent Hence, the required density resulting in the field equation with
the derivatives up to second order has the following form:

\begin{equation}
L_{g}=-\omega _{1}\widetilde{g}^{\mu \nu }\left( F_{\mu \nu }^{\lambda
}F_{\lambda \kappa }^{\kappa }-F_{\mu \kappa }^{\lambda }F_{\nu \lambda
}^{\kappa }\right) +\omega _{2}\sqrt{-g},  \label{eq6}
\end{equation}

\noindent where $\omega _{1}$ and $\omega _{2}$ are the some constants.\

However, the structure of the Lagrangian density (\ref{eq6})\ does not allow
including the background metrics in the field equations. Therefore we have
to add in Eq. (\ref{eq6}) the terms explicitly containing $\gamma _{\mu \nu }
$\ and violating the gauge group under consideration \cite{logunov3,logunov4}%
. The term $\gamma _{\mu \nu }\widetilde{g}^{\mu \nu }$\ obeys the required
transformational properties but only for the gauge vectors:

\begin{equation}
g^{\mu \nu }D_{\mu }D_{\nu }\zeta ^{\lambda }=0.  \label{eq7}
\end{equation}

\noindent\ Resulting Lagrangian density for the gravitational field is:

\begin{equation}
L_{g}{\acute{}}=-\omega _{1}\widetilde{g}^{\mu \nu }\left( F_{\mu \nu
}^{\lambda }F_{\lambda \kappa }^{\kappa }-F_{\mu \kappa }^{\lambda }F_{\nu
\lambda }^{\kappa }\right) +\omega _{2}\sqrt{-g}+\omega _{3}\gamma _{\mu \nu
}\widetilde{g}^{\mu \nu }+\omega _{4}\sqrt{-\gamma },  \label{eq8}
\end{equation}

\noindent here the last term is introduced to provide the vacuum stability,
i.e. to exclude the cosmological constant-like term in the absence of the
matter.

From the variational principle for the gravitational field ($\frac{\delta
L_{g}{\acute{}}}{\delta \widetilde{g}^{\mu \nu }}=0$), the requirement of
vacuum stability and taking into account the material sources for the
gravitational field we can obtain from (\ref{eq8}) the field equation:

\begin{equation}
G_{\nu }^{\mu }-\frac{m^{2}}{2}\left( \delta _{\nu }^{\mu }+g^{\mu \lambda
}\gamma _{\lambda \nu }-\frac{1}{2}\delta _{\nu }^{\mu }g^{\kappa \lambda
}\gamma _{\kappa \lambda }\right) =-\frac{8\pi \varkappa }{c^{4}}T_{\nu
}^{\mu },  \label{eq9}
\end{equation}

\noindent where $m^{2}=\left( m_{g}c\diagup \hbar \right) ^{2}$, $\varkappa $%
\ is the Newtonian gravitational constant, $m_{g}$ is the graviton mass as a
natural interpretation of the constants $\omega $ incoming in the Lagrangian
density, $G_{\nu }^{\mu }$ is the Einstein tensor. Below we shall use $%
c=\hbar =1$, then $m_{pl}=1\diagup \sqrt{8\pi \varkappa }=2.43\times
10^{18}\ GeV$ is the reduced Planck mass. It should be noted, that the mass
of graviton results from the gauge group violation, i.e. it appears together
with the background metrics in the Lagrangian. Otherwise we have the usual
Einstein-Gilbert field equations (without cosmological constant) and the
background spacetime loses its physical meaning.

Now let us consider the physical sense of the constraint (\ref{eq7}). As a
matter of fact, the introduced simplest modification of the Lagrangian by
the term $\gamma _{\mu \nu }\widetilde{g}^{\mu \nu }$ violating the gauge
group results in the equation:

\begin{equation}
D_{\mu }\widetilde{g}^{\mu \nu }=0,  \label{eq10}
\end{equation}

\noindent which is the consequence of the field equations and defines the
polarization of the gravitational field (spin states 2 and 0) \cite{logunov4}%
. So, the structure of the mass part in the field equations and the field
polarization are interdependent.

Now we have to consider an important consequence of the considered bi-metric
approach. The point is that the existence of the physically meaningful
background spacetime imposes the \textit{causality principle}, which
constraints the permissible solutions in the RTG. This background defines
the observable events and the corresponding relations between them. These
relations always can be attributed to the Minkowski spacetime. Hence, the
causality cone of the effective Riemannian spacetime should be positioned
inside the causality cone of the Minkowski spacetime \cite{logunov5}:

\begin{gather}
\gamma _{\mu \nu }u^{\mu }u^{\nu }=0,  \label{eq11} \\
g_{\mu \nu }u^{\mu }u^{\nu }\leq 0,  \notag
\end{gather}

\noindent where $u^{\mu }$ is the arbitrary isotropic vector.

The cosmological equations in the RTG can be obtained on the general basis.
However, we have to take into account that the formally arbitrary choice of
the convenient $g^{\mu \nu }$, which is the typical trick in the GR, is not
always appropriate in the RTG because this implies the simultaneous
constraints on $\gamma _{\mu \nu }$.

Let us consider the homogeneous and isotropic Riemannian spacetime induced
by the global gravitational field. As it was above mentioned, this spacetime
in the framework of the RTG is \textit{flat}. This is the consequence of the
field equations (see \cite{logunov2,logunov3,kalash1}). The corresponding
interval in the spherical coordinates is \cite{logunov3}:

\begin{equation}
ds^{2}=d\tau ^{2}-\alpha a\left( \tau \right) \left[ dr^{2}+r^{2}\left(
d\theta ^{2}+\sin \left( \theta \right) ^{2}d\phi ^{2}\right) \right] ,
\label{eq12}
\end{equation}

\noindent where $\tau $\ is the proper time, $a\left( \tau \right) $ is the
scaling factor and $\alpha $\ is the constant of integration (its meaning
see below).

Let's the background is described by the Galilean metrics. Eqs. (\ref{eq12}, %
\ref{eq11}) result in \cite{logunov3}:

\begin{equation}
a\left( \tau \right) ^{4}-\alpha <0,  \label{eq13}
\end{equation}

\noindent which eliminates the cosmological solution with the eternal
expansion. This is the consequence of the causality principle in the RTG. It
is convenient to assign $\alpha =a_{\max }^{4}$, where $a_{\max }$ is the
maximum scaling factor.

Then the cosmological equations are:

\begin{equation}
\left( \frac{\overset{\text{\textperiodcentered }}{a}}{a}\right) ^{2}=\frac{%
\rho \left( \tau \right) }{3m_{pl}^{2}}-\frac{m^{2}}{12}\left( 2+\frac{1}{%
a\left( \tau \right) ^{6}}-\frac{3}{a\left( \tau \right) ^{2}a_{\max }^{4}}%
\right) ,  \label{eq14}
\end{equation}

\begin{equation}
\frac{\overset{\text{\textperiodcentered \textperiodcentered }}{a}}{a}=-%
\frac{\rho \left( \tau \right) +3p\left( \tau \right) }{6m_{pl}^{2}}-\frac{%
m^{2}}{6}\left( 1-\frac{1}{a\left( \tau \right) ^{6}}\right) ,  \label{eq15}
\end{equation}

\noindent and $a\left( \tau \right) \leq a_{\max }$. $\rho $ and $p$ are the
matter density and pressure, respectively; the dot denotes the derivative
with respect to $\tau $. These equations are similar to those in the GR with
the flat global spacetime but: 1) they contain the terms describing the
massive graviton and 2) suppose the increase of $a$ up to some maximum
scaling factor $a_{\max }$ as a result of the causality principle.

\section{Cosmological scenarios in the RTG and constraints on the graviton
mass}

Before an examination of the cosmological scenarios, let us consider
possible embedding of the effective Riemannian spacetime in the background
with the constant curvature and the same dimension. The hyperbolic
background has to be rejected due to the causality principle violation. The
causally connected events in the effective spacetime are asymptotic causally
independent on the background.

\begin{equation}
a^{4}\leq \frac{\alpha }{1+r^{2}}\underset{r\rightarrow \infty }{%
\longrightarrow }0.  \label{eq16}
\end{equation}

However, the spherical background obeys the causality principle. The
corresponding global Riemannian spacetime is spherical, too. Then the
cosmological equations have the modified form:

\begin{equation}
\left( \frac{\overset{\text{\textperiodcentered }}{a}}{a}\right) ^{2}=\frac{%
\rho }{3m_{pl}^{2}}-\frac{m^{2}}{12}\left( 2+\frac{1}{a^{6}}-\frac{3}{%
a^{2}a_{\max }^{4}}\right) -\frac{1}{a^{2}a_{\max }^{4}},  \label{eq17}
\end{equation}

\begin{equation}
\frac{\overset{\text{\textperiodcentered \textperiodcentered }}{a}}{a}=-%
\frac{\rho +3p}{6m_{pl}^{2}}-\frac{m^{2}}{6}\left( 1-\frac{1}{a^{6}}+\frac{%
3\left( 1-\Sigma ^{2}\right) }{4a^{2}a_{\max }^{4}}\right) ,  \label{eq18}
\end{equation}

\noindent where $\Sigma $ is the background curvature, $a_{\max }=\Sigma $.
\

Turning back we can conclude that, although it is possible to embed the
effective spacetime into the spherical background, there are no some
physical justifications for such complication of the model. Nevertheless,
the extension of the background dimensionality requires an additional
analysis but this exceeds the limits of this article \cite{kaluza}.

Let us return to Eqs. (\ref{eq14}, \ref{eq15}) and consider their structure.
It is clear that the fulfilment of the causality principle requiring only
closed evolutional scenarios results from the first term in the brackets of
Eq. (\ref{eq14}). This term produced by the massive graviton plays a role of
the negative cosmological constant, which stops any expansion of the
universe with the arbitrary material contents if the state parameter for
their dominating form is $w>-1$. The corresponding minimum density is
connected with the graviton mass and the maximum scaling factor \cite%
{logunov3}:

\begin{equation}
\rho _{\min }=\frac{m^{2}m_{pl}^{2}}{2}\left( 1-\frac{1}{a_{\max }^{6}}%
\right) .  \label{eq19}
\end{equation}

On the other hand, when the scaling factor is small, the second term in the
brackets of Eq. (\ref{eq15}) causes the repulsion (antigravitation) induced
by the graviton mass in the strong gravitational field. This repulsion
prevents from the existence of the initial cosmological singularity and
provides the acceleration at the initial stage of the universe expansion.
However, as one can see from Eq. (\ref{eq15}), out of this initial stage
there is only decelerated expansion up to $a_{\max }$ if the state parameter
for the dominating form of the matter obeys $w>-\frac{1}{3}$. We remind,
that as a result of the vacuum stability principle (i.e. due to $g_{\mu \nu }%
\underset{T_{\mu \nu }\longrightarrow 0}{\longrightarrow }\gamma _{\mu \nu }$%
) the cosmological constant in the RTG has only gravitational nature and its
sign is negative (i.e. it causes the attraction on the large scales). Here
we face the challenge of the disagreement with the modern observational data.

The data obtained from the BOOMERANG, MAXIMA and COBE projects \cite%
{riess,permutter,jaffe} suggest the accelerated expansion of the universe at
present. The acceleration parameter can be estimated as $q\equiv \left(
d^{2}a\diagup d\tau ^{2}\right) \mid _{0}\diagup \left(
a_{0}H_{0}^{2}\right) $ $\simeq 0.33\pm 0.17$ and has a positive value (here
$H$ is the Habble constant and the zero index refers to the present epoch
when $\tau =\tau _{0}$). On the whole the data are summarized in Table \ref%
{table} (see also \cite{krauss}).

\begin{table}[tbp]
\caption{Cosmological parameters}
\label{table}\centering%
\begin{tabular}{|c|c|}
\hline
\textbf{Cosmological parameters} & \textbf{Observational data} \\ \hline
$H_{0},~km\diagup \left( s\cdot Mps\right) $ & $68\pm 6$ \\ \hline
$\Omega _{tot}$ & $1.11\pm 0.07$ \\ \hline
$\Omega _{m}$ & $0.37\pm 0.07$ \\ \hline
$\Omega _{r}$ & $\left( 9.34\pm 1.64\right) \times 10^{-5}$ \\ \hline
$\Omega _{x}$ & $0.71\pm 0.05$ \\ \hline
$\tau _{0},~Gyr$ & $12.7\pm 3$ \\ \hline
$w$ & $\leq -0.6$ \\ \hline
$q$ & $0.33\pm 0.17$ \\ \hline
\end{tabular}%
\end{table}

The age of the universe $\tau _{0}$ is estimated from the age of the oldest
globular clusters. $\Omega _{m}\equiv \rho _{m}\diagup \left(
3m_{pl}^{2}H_{0}^{2}\right) $ is the density parameter for the
\textquotedblleft normal\textquotedblright\ matter at present. The word
\textquotedblleft normal\textquotedblright\ means that this matter possesses
the state parameter obeying $w\geq 0$. Such matter can only decelerate the
cosmological expansion. However, the detailed structure of this
\textquotedblleft normal\textquotedblright\ sector of the matter is unknown.
The baryons contribution amounts only $\simeq 5\%$ in the total density and
the rest of the matter sector belongs to the so-called cold dark matter,
which is not revealed to day.

The similar parameter for the photons is $\Omega _{\gamma }\equiv \rho
_{\gamma }\diagup \left( 3m_{pl}^{2}H_{0}^{2}\right) =2.51\times
10^{-5}h^{-2}\simeq \left( 5.56\pm 0.97\right) \times 10^{-5}$ (here $%
H_{0}\equiv 100\times h\ km\diagup \left( s\cdot Mps\right) $\ ), and for
the massless neutrino $\Omega _{\nu }=0.681\times \Omega _{\gamma }$ \cite%
{lyth}. Then for the relativistic matter (\textquotedblleft
radiation\textquotedblright ) we have $\Omega _{r}\simeq \left( 9.34\pm
1.64\right) \times 10^{-5}$.

The accelerated expansion of the universe suggests that the main part of the
density in the universe belongs to some exotic \textquotedblleft dark
energy\textquotedblright\ or \textquotedblleft X-matter\textquotedblright\
with the density parameter $\Omega _{x}$ and the state parameter obeying $%
w\leq -0.6$. The latter provides the negative pressure and, as a result, the
acceleration of the universe expansion.

The parameter $\Omega _{tot}\equiv \Omega _{x}+\Omega _{m}+\Omega _{r}$\
defines the curvature of the effective Riemannian spacetime in the GR
through the so-called cosmic sum rule: $\Omega _{tot}+\Omega _{K}=1$, where $%
\Omega _{K}\equiv -K\diagup \left( a_{0}^{2}H_{0}^{2}\right) $ and $K=1,~-1$
and $0$ for the spherical, hyperbolic and flat spacetimes, respectively. As $%
\Omega _{tot}\simeq 1$ at present \cite{bernardis}, this requires the fine
tuning at past (for example, if we start from the Planck scale the deviation
from the unity at the beginning of the expansion is about of $10^{-60}$ \cite%
{peacock}). There is no such problem in the RTG because the flatness of the
homogeneous and isotropic Riemannian spacetime is the consequence of the
field equations.

Now let us consider the possible modifications of the cosmological scenarios
in the framework of the RTG, which provide the agreement with the modern
observational data. To obtain the accelerated expansion at present we modify
the energy-momentum tensor by the insertion of the quintessence term with
the negative pressure. The practically interesting candidate for the
quintessence is some scalar field $\phi $, which evaluates slowly in a
runaway potential $V\left( \phi \right) $: $V\left( \phi \right) \underset{%
\phi \longrightarrow \infty }{\longrightarrow }0$ \cite%
{wetterich,ratra,caldwell}.

In the beginning let's consider the problem phenomenologically and introduce
the quintessential term with the constant state parameter $%
w_{x}=p_{x}\diagup \rho _{x}$ lying within the limits of the strong and week
energy conditions: $-1<w<-1\diagup 3$ \cite{kalash2,kalash3}. It is
convenient to suppose that at present $a\left( \tau _{0}\right) =1$ and to
transit from the densities to the density parameters. Then Eq. (\ref{eq14})
can be rewritten as:

\begin{gather}
H\left( \tau \right) ^{2}=H_{0}^{2}\times  \label{eq20} \\
\left[ \frac{\Omega _{r}}{a\left( \tau \right) ^{4}}+\frac{\Omega _{m}}{%
a\left( \tau \right) ^{3}}+\frac{\Omega _{x}}{a\left( \tau \right) ^{\left(
3+3w_{x}\right) }}-\Omega _{g}\left( 1+\frac{1}{2a\left( \tau \right) ^{6}}-%
\frac{3}{2a\left( \tau \right) ^{2}a_{\max }^{4}}\right) \right] ,  \notag
\end{gather}

\noindent where we used $\rho \left( \tau \right) \propto a\left( \tau
\right) ^{-3\left( 1+w\right) }$ and introduced $\Omega _{g}=m^{2}\diagup
\left( 6H_{0}^{2}\right) $ (the density parameter for the massive graviton).

We can see from Eq. (\ref{eq20}) that the massive graviton modifies the
cosmic sum rule

\begin{equation}
\Omega _{r}+\Omega _{m}+\Omega _{x}-\frac{3}{2}\Omega _{g}\left( 1-\frac{1}{%
a_{\max }^{4}}\right) =1,  \label{eq21}
\end{equation}

\noindent which is like that for the spherical curvature of the effective
Riemannian spacetime ($a_{\max }\gg 1$). This similarity results from the
negative cosmological constant-like action of the gravitons. Note however
that in fact the spacetime is flat.

The substitution $t=H_{0}\left( \tau -\tau _{0}\right) $\ in Eq. (\ref{eq15}%
) produces (this substitution really omits $H_{0}^{2}$ from the right-hand
side of Eq. (\ref{eq20})):

\begin{equation}
\frac{d^{2}a\left( t\right) }{dt^{2}}=-\frac{\Omega _{m}}{2a\left( t\right)
^{2}}-\frac{\Omega _{r}}{a\left( t\right) ^{3}}-\frac{1+3w_{x}}{2a\left(
t\right) ^{\left( 2+3w_{x}\right) }}-\Omega _{g}\left( a\left( t\right) -%
\frac{1}{a\left( t\right) ^{5}}\right) .  \label{eq22}
\end{equation}

Eqs. (\ref{eq20}, \ref{eq22}) result in the expression for the acceleration
parameter:

\begin{equation}
q=\frac{\Omega _{x}\left( 1-\frac{3}{2}\chi \right) -\frac{1}{2}\Omega
_{m}-\Omega _{r}}{\Omega _{tot}-\frac{3}{2}\Omega _{g}},  \label{eq23}
\end{equation}

\noindent where $\chi \equiv 1+w_{x}$ is the deviation of the quintessence
state parameter from that for the pure positive cosmological constant. If
the gravitons and the relativistic matter do not contribute to the present
state, the combination of the observational data and Eq. (\ref{eq23})
results in the estimation for $\chi $:

\begin{equation}
\chi =\frac{2}{3}\left( 1-q\right) -\frac{\Omega _{m}}{3\Omega _{x}}\left(
1+2q\right) \simeq 0.16_{-0.09}^{+0.11}.  \label{eq24}
\end{equation}

\noindent The deviation of the state parameter from that for the pure
cosmological constant can be considered as the justification of the initial
guess about the material (not vacuum) source of the accelerated expansion.

If $a_{\max }\gg a_{0}$ (this is a well-grounded assumption because the
graviton mass has to be small, see below), then the minimum density is
defined by the material terms with a slowest density decrease due to the
scaling factor increase. These are the negative cosmological constant
produced by the massive graviton and the quintessence with small $\chi $.
Hence we have the estimation for the maximum scaling factor:

\begin{equation}
\frac{\Omega _{g}}{\Omega _{x}}\simeq a_{\max }^{-3\chi }.  \label{eq25}
\end{equation}

\noindent As a result, Eqs. (\ref{eq24}, \ref{eq25}) give the dependence of
the maximum scaling factor on the graviton density parameter. It is natural,
the approach of $w_{x}$\ to $-1$ and $\Omega _{x}$\ to $1$ increase the
maximum scaling factor due to growing negative pressure of the quintessence.

Eqs. (\ref{eq20}, \ref{eq25}) allow finding the minimum scaling factor. The
corresponding equation is:

\begin{equation}
\Omega _{g}a^{3w_{x}}\left( -2a^{6}+\frac{3}{a_{\max }^{4}}a^{4}-1\right)
+2a^{3w_{x}}\left( \Omega _{r}a^{2}+\Omega _{m}\right) +2\Omega _{x}a^{3}=0.
\label{eq26}
\end{equation}

\noindent If $w_{x}\gtrapprox -1$ and $a_{\max }\gg 1$ then $a_{\min }\simeq
\sqrt{\Omega _{g}\diagup \left( 2\Omega _{r}\right) }$. \ It is obviously
that the minimum scaling factor can not be less than that corresponding to
the radiation domination epoch. For the radiation domination we have the
well-known condition (if the universe thermalized):

\begin{equation}
\rho _{\max }=\frac{\pi ^{2}}{30}g_{\ast }\left( T\right) T_{\max }^{4},
\label{eq27}
\end{equation}

\noindent where $g_{\ast }(T)$ is the effective degeneracy factor, $T$ is
the temperature. Simultaneously, as the scaling factor is roughly $a_{\min
}\simeq T_{0}/T_{\max }$ ($T_{0}\simeq 10^{-4}~eV$ is the present
temperature of the cosmic background), we obtain the expression for the
graviton density and the maximum scaling factor:

\begin{eqnarray}
\Omega _{g} &\simeq &2\Omega _{r}\left( \frac{T_{0}}{T_{\max }}\right) ^{2},
\label{eq28} \\
a_{\max } &\simeq &\sqrt[3\varkappa ]{\left( \frac{\Omega _{x}}{2\Omega _{r}}%
\right) \left( \frac{T_{\max }}{T_{0}}\right) ^{2}}.  \notag
\end{eqnarray}

The maximum admissible graviton mass and the maximum scaling factor are
presented in Table \ref{table2} (for the cosmological parameters we choose
their mean values). The estimation of the maximum graviton mass means that
the universe starts its expansion from the denoted \textquotedblleft
event\textquotedblright\ (the dimensional mass can be re-calculated by means
of the relation $m_{g}=\sqrt{6\Omega _{g}}H_{0}\hbar \diagup c^{2}$).

\begin{table}[tbp]
\caption{Estimations for the maximum graviton mass and the maximum scaling
factor. GUT is the grand unified theory phase transition, EW is the
electroweak phase transition, NS is the nucleosynthesis, RD is the end of
the radiation domination}\centering%
\begin{tabular}{|l|l|l|l|}
\hline
\textbf{Event} & $\mathbf{T}$ & $\mathbf{m}_{g},~g$ & $\mathbf{a}_{\max }$
\\ \hline
GUT & $\sim 10^{15}~GeV$ & $\sim 10^{-94}~\left( \sim 10^{-61}~eV\right) $ &
$10^{111}$ \\ \hline
EW & $\sim 100~GeV$ & $\sim 10^{-82}~\left( \sim 10^{-49}~eV\right) $ & $%
10^{53}$ \\ \hline
NS & $\sim 0.1~MeV$ & $\sim 10^{-76}~\left( \sim 10^{-43}~eV\right) $ & $%
10^{28}$ \\ \hline
RD & $\sim 1~eV$ & $\sim 10^{-71}~\left( \sim 10^{-38}~eV\right) $ & $10^{7}$
\\ \hline
\end{tabular}%
\label{table2}
\end{table}

As it was above mentioned, the RTG solves some basic problems, which inspire
the inflation paradigm in the modern cosmology: the flatness problem and the
problem of the source of the initial expansion. Moreover, the inflation does
not solve the problem of the singularity \cite{vilenkin1}, which is lacking
in the RTG. The problem of the relics in the RTG can be solved if $a_{\min }$
is too large to provide $T_{\max }$, which is sufficient for their creation.
However, the problem of the horizon remains: the size of the causally
connected domains at the moment of the last scattering of the cosmic
background photons is $\sim $100 Mps. In principle, this problem can be
solved without inflation (see, for example \cite{durrer}). As the RTG
eliminates singularity, it admits the physically meaningful oscillating
solution with increasing homogeneity and isotropy. Nevertheless, let us
examine the compatibility of the RTG with the inflation paradigm.

As it was mentioned, the feature of the RTG is the antigravitation produced
by the massive graviton in the strong gravitational field. This causes the
accelerated expansion at the initial stage of the universe evolution and
prevents from the singularity. However, the gravitational field is produced
by the matter therefore the character of the initial acceleration is defined
by the form of this matter. As an example, the relativist matter (radiation)
results in the short acceleration stage (\textquotedblleft
inflation\textquotedblright ) $t_{ac}=\Omega _{g}^{3\diagup 2}\left[
1-1\diagup \sqrt{2}\right] \diagup \left( 3\Omega _{r}^{5\diagup 2}\right) $
($t_{ac}$ is the acceleration time) with the scaling factor growing by only
factor of root of two: $a_{end}\diagup a_{\min }=\sqrt{2}$. It is clear that
such short inflation is not sufficient for the solution of the horizon
problem.

The appropriate choice is the inflation governed by the scalar field $\phi $
(inflaton). Let's consider the minimally coupled single scalar field with
the potential $V\left( \phi \right) $. Then the initial evolution can be
described by the following system:

\begin{gather}
\left( \frac{\overset{\text{\textperiodcentered }}{a}}{a}\right) ^{2}=\frac{1%
}{3m_{pl}^{2}}\left( \frac{\overset{\text{\textperiodcentered }}{\phi }}{2}%
+V\left( \phi \right) \right) -\frac{m^{2}}{12a^{6}},  \label{eq29} \\
\overset{\text{\textperiodcentered \textperiodcentered }}{\phi }+3\phi \frac{%
\overset{\text{\textperiodcentered }}{a}}{a}=-\frac{dV}{d\phi }.  \notag
\end{gather}

\noindent Note, that at the beginning there exist no other material fields
with the exception of the scalar field. If at the beginning the potential
energy prevails over the kinetic one, the exponential expansion begins (the
standard slowroll conditions have to be satisfied: $m_{pl}\left\vert \frac{dV%
}{d\phi }\diagup V\right\vert \ll 1$ and $m_{pl}^{2}\left\vert \frac{d^{2}V}{%
d\phi ^{2}}\diagup V\right\vert \ll 1$). The simple estimation shows that
the graviton term vanishes very quickly and the expansion does not differ
from that in the GR. The inflation ends when the sufficient energy transfers
into the kinetic form. Then the so-called reheating begins and the material
fields are created. The natural criterion providing this primordial
inflation in the framework of the RTG is:

\begin{equation}
a_{\min }\leq a_{begin}\ll a_{end},  \label{eq30}
\end{equation}

\noindent where $a_{begin}$ and $a_{end}$ are the scaling factors at the
beginning and at the end of the inflation, respectively.

Let us consider the potential, which admits the primordial inflation solving
the horizon problem and, simultaneously, allows the second inflation
describing the accelerated expansion at present. Such models consider both
inflations as the manifestation of the single scalar field (quintessential
inflation models, for review see \cite{dimopoulos}). For example, the
potential \cite{vilenkin2}

\begin{eqnarray}
V &=&\lambda \left( \phi ^{4}+M^{4}\right) \text{ for}~\phi <0,  \label{eq31}
\\
&=&\frac{\lambda M^{8}}{\phi ^{4}+M^{4}}\text{ \ \ \ \ \ for }\phi \geq 0
\notag
\end{eqnarray}

\noindent corresponds to the case of the self-interacting $\lambda \phi
^{4}- $ field for the negative $\phi $\ (the value of the cosmic background
fluctuations requires $\lambda \leq 10^{-14}$) and provides the second
acceleration on the rolling-away tail of the potential, when $\phi
\longrightarrow \infty $ (quintessential tail). The present value of $\Omega
_{x}$ requires $M\simeq 8\times 10^{5}~GeV$.

The first inflation terminates at $\left\vert \phi \right\vert \sim m_{pl}$
and Eqs. (\ref{eq29}, \ref{eq30}) result in ($M\ll m_{pl}$):

\begin{equation}
a_{end}\gg \sqrt[3]{\frac{m}{m_{pl}\sqrt{\lambda }}}.  \label{eq32}
\end{equation}

\noindent Simultaneously,

\begin{equation}
\frac{a_{r}}{a_{end}}\simeq \frac{1}{\sqrt{\lambda R}},  \label{eq33}
\end{equation}

\noindent where $R\simeq 0.01$ is the numerical factor defining the
particles creation at the end of the first inflation and the transit to the
radiation domination \cite{vilenkin2}.\ At the beginning of the radiation
domination, when $a\equiv a_{r}$, the temperature was

\begin{equation}
T_{r}\simeq \lambda R^{3\diagup 4}m_{pl}\simeq 10^{3}~GeV.  \label{eq34}
\end{equation}

\noindent Then roughly we have:

\begin{equation}
m\ll \lambda ^{2}R^{3\diagup 2}m_{pl}\left( \frac{T_{0}}{T_{r}}\right)
^{3}\simeq 10^{-52}~eV,  \label{eq35}
\end{equation}

\noindent If we made the usual assumption about the $60$-$e$ folding
expansion during the inflation then $a_{begin}\simeq a_{end}\times e^{-60}$.
Hence we have the estimation for the graviton mass allowing the appropriate
scaling factor:

\begin{gather}
m\sim \frac{\sqrt{12\lambda }}{m_{pl}}\phi _{in}^{2}\left( a_{end}\times
e^{-60}\right) ^{3}\simeq \frac{\sqrt{12}}{m_{pl}}\lambda ^{2}\Pi ^{3\diagup
2}\phi _{in}^{2}\left( \frac{T_{0}}{T_{r}}\right) ^{3}\times e^{-180}
\label{eq36} \\
\simeq 10^{-157}\frac{\phi _{in}^{2}}{m_{pl}},  \notag
\end{gather}

\noindent where $\phi _{in}$ (initial scalar field) and graviton mass are
expressed through the Planck mass. Although $\left\vert \phi
_{in}\right\vert \gg m_{pl}$, the obtained estimation is extremely low
because it is very hard to \textquotedblleft squeeze\textquotedblright\ the
universe down to the Planck scale in the condition of the strong
antigravitation produced by the massive graviton. Such low value for the
graviton mass can not be attributed to some real physics. However, we have
not to consider this conclusion as the pessimistic estimation of the
incompatibility between the RTG and the primordial inflation picture because
1) our estimation is the model-dependent and needs an additional
investigation; 2) we have not to overestimate our knowledge of the physics
on the Planck scale; 3) the RTG can propose an alternative (oscillating)
scenario without primordial inflation.

It is of interest to consider the compatibility of the RTG with the second
inflation picture, which takes a place on the quintessential tail of the
model under consideration. In the framework of this model (see Eq. (\ref%
{eq31}) (the below described picture is common for the different models of
the quintessential inflation, see \cite{dimopoulos}) we have the following
evolutional stages: 1) \emph{First inflation}. The field $\phi \ll -m_{pl}$
slowly rolls to zero. The potential energy dominates over the kinetic one.
As a consequence, the state parameter $w_{x}=\frac{\overset{\text{%
\textperiodcentered }}{\phi }^{2}\diagup 2-V}{\overset{\text{%
\textperiodcentered }}{\phi }^{2}\diagup 2+V}\approx -1$. 2) \emph{Reheating
and kination}. $\phi >-m_{pl}$ causes the end of the inflation due to the
kinetic term increase ($w_{x}\longrightarrow 1$). The energy transfers to
the material fields. But the kinetic dominated scalar field decreases as $%
\rho _{x}\approx \frac{\overset{\text{\textperiodcentered }}{\phi }^{2}}{2}%
\propto a^{-6}$. Hence, the radiation (and then the matter) domination
begins. 3) The kinetic term vanishes and the potential energy of the scalar
field dominates again. \emph{Second inflation} begins from which the
universe never recovers because the slowroll conditions are satisfied.

However, the RTG provides the quite natural exit from the eternal inflation
due to the presence of the negative cosmological constant-like term in Eq. (%
\ref{eq17}). As $V$ decreases as \cite{vilenkin2}

\begin{equation}
V\sim \frac{\lambda M^{8}}{m_{pl}^{2}\ln ^{4}\left( \frac{a}{a_{end}}\right)
},  \label{eq37}
\end{equation}

\noindent the inflation stops when

\begin{equation}
a\simeq a_{end}\exp \left( \frac{M^{2}\sqrt{mm_{pl}\sqrt{\lambda }}}{%
mm_{pl}^{2}}\right) \sim 10^{-24}\exp \left( \frac{10^{-14}}{\sqrt{m\left[ eV%
\right] }}\right) .  \label{eq38}
\end{equation}

\noindent From Eq. (\ref{eq38}) the maximum mass of the graviton is $\simeq
10^{-31}~eV$ (criterion $a_{\max }>1$), however, the maximum scaling factor
increases exponentially with the $m$ decrease in contrast to Eq. (\ref{eq25}%
) because $w_{x}\approx -1$ in the late universe. Here we do not consider
the additional numerical estimations because they are model-dependent.
Nevertheless, it is obviously that the combination of the first inflation
condition ($a_{\min }\ll a_{end}$) with the break of second inflation can
result in the exponentially large $a_{\max }$.

\section{Conclusions}

The RTG is able to solve some important cosmological problems. It does not
contain the cosmological singularity and derives the flatness of the global
spacetime from the field equations. The antigravitation produced by the
massive graviton in the strong gravitational field solves the problem of the
source of the initial expansion and allows escaping the relics creation.
However, the problems of the horizon and the present accelerated expansion
of the universe remain. The former is solvable in the framework of the
oscillation paradigm. The RTG admits only closed evolutional scenario in
virtue of the causality principle and thereby the causal connections with
the extremely distant domains can result from the previous cycles of the
oscillation (remind that there is no the singularity in the RTG). However,
the problem of the accelerated expansion needs some additional hypothesis.
The appropriate modification of the RTG Lagrangian is awkward because the
structure of its massive part is defined by the polarization properties of
the gravitational field. The alternative way is the modification of the
energy-momentum tensor due to inclusion of the so-called quintessence term
with the state parameter lying within the limits of the strong and weak
energy conditions that causes the repulsion and, as a result, the
accelerated expansion.

In the framework of the latter approach there is the single scenario:

\begin{enumerate}
\item \emph{First acceleration (inflation)}, which can be governed by the
scalar field (exponential inflation) or by the massive gravitons (power-mode
inflation that occurs if the radiation dominates at the beginning of the
expansion).

\item \emph{First deceleration }due to the radiation (and then matter)
domination. The massive gravitons do not contribute due to the increased
scaling factor.

\item \emph{Second acceleration (inflation)} due to the quintessence
domination. The massive gravitons do not contribute.

\item \emph{Second deceleration} due to the negative cosmological
constant-like action of the massive gravitons.

\item \emph{Contraction} produced by the massive gravitons.
\end{enumerate}

\noindent At the first stage, there are the certain constraints on the
graviton mass: the initial scaling factor has to provide the temperature,
which is sufficient for the formation of the universe in its known form. At
least, this temperature has to exceed that required for the nucleosynthesis.
As a result, $m_{g}<10^{-43}~eV$. These constraints become extremely
exacting in the case of the primordial inflation governed by the scalar
field because it is hard to squeeze the universe down to the Planck size.
One could say that the RTG is hardly compatible with such primordial
inflation.

The second inflation can be considered in two ways. Firstly, we can suppose
the constant state parameter for the quintessence: $w_{x}>-1$. In this case
the massive graviton terminates the inflation for the scaling factor, which
is power dependent on the graviton mass. The rough estimation for the above
given $m_{g}$ results in the relative scaling factor $\sim 10^{28}$ (if its
present value is $1$). Secondly, we can consider the quintessence as some
scalar field with the rolling-away potential. This is a more complicated
situation because the state parameter changes and approaches $-1$ in the
late universe. However, the massive graviton stops the inflation in this
case too, but the dependence of the maximum scaling factor on the graviton
mass is exponential. As the latter approach is based on the artificial model
building, the problem of the late universe evolution in the framework of the
RTG needs an additional investigation.

In spite of the successful agreement of the RTG with the modern
observational data, the unsolved problems remain:

\begin{enumerate}
\item The horizon problem and the initial expansion of the universe remain
unexplored in the RTG. There are some doubts about the compatibility of the
RTG with the primordial inflation governed by the scalar field.

\item The quintessential scenarios need a more detailed investigation.
Moreover, the nature of the quintessence is still unknown and this
hypothesis faces some typical problems:

\begin{enumerate}
\item the quintessence has to be extremely weakly coupled with the usual
matter;

\item it is probably that the quintessence has to be a very light \cite%
{carroll};

\item the quintessence would generate the corrections to the gauge coupling.
\end{enumerate}

\item And at last, why is the graviton so light? It is necessary to explore
the connections of the RTG with the modern field theory.
\end{enumerate}

\section*{\protect\bigskip Acknowledgements}

Author appreciates Prof. A. A. Logunov and Prof. M. A. Mestvirishvili for
the promotional and helpful discussions and Organizing Committee for the
hearty welcome, the creative and friendly atmosphere at Workshop. Author is
Lise Meitner Fellow, supported by the Austrian National Science Found (grant
M688).

\end{document}